\documentclass[12pt]{iopart}

\usepackage{amssymb}
\usepackage{graphicx}
\usepackage{bm}

\begin{document}
\title{Evolving networks and the development of neural systems}

\author{Samuel Johnson, J. Marro and Joaqu{\'i}n J. Torres}

\address{Departamento de Electromagnetismo y F{\'i}sica de la Materia and Instituto Carlos I de F{\'i}sica Te{\'o}rica y Computacional, Facultad de Ciencias, Universidad de Granada, 18071 Granada, Spain.\\
E-mail: samuel@onsager.ugr.es, jmarro@ugr.es and jtorres@onsager.ugr.es
}

\begin{abstract}
It is now generally assumed that the heterogeneity of most networks in nature probably arises via preferential attachment of some sort. However, the origin of various other topological features, such as degree-degree correlations and related characteristics, is often not clear and attributed to specific functional requirements. We show how it is possible to analyse a very general scenario in which nodes gain or lose edges according to any (e.g., nonlinear) functions of local and/or global degree information. Applying our method to two rather different examples of brain development -- synaptic pruning in humans and the neural network of the worm {\it C. Elegans} -- we find that simple biologically motivated assumptions lead to very good agreement with experimental data. In particular, many nontrivial topological features of the worm's brain arise naturally at a critical point.

\end{abstract}

\submitto{J. Stat. Mech. (accepted, 2010)}
\maketitle

\section{Introduction}

The conceptual simplicity of a \textit{network} -- a set of nodes, some pairs of which connected by edges -- often suffices to capture the essence of cooperation in complex systems. Ever since Barab\'{a}si and Albert presented their evolving network model \cite{Barabasi}, in which linear preferential attachment leads asymptotically to a scale-free degree distribution (the degree, $k$, of a node being its number of neighbouring nodes), there have
been many variations or refinements to the original scenario
\cite{Albert,Bianconi_bose,Krapivsky,Bianconi_fitness,Park_Li,Suhan} (for a review, see \cite{Boccaletti}).
In \cite{Johnson_PRE}, we showed how topological phase transitions and scale-free solutions could emerge in the case of nonlinear rewiring in fixed-size networks. Now we extend our scope to more general and realistic situations, considering the evolution of networks
making only minimal assumptions about the attachment/detachment rules. In fact, all we assume is that these probabilities factorize into two parts: a local term which
depends on node degree, and a global term, which is a function of the
mean degree of the network.

Our motivation can be found in the
mechanisms behind many real-world networks, but we focus, for the sake of
illustration, on the development of biological neural networks,
where nodes
represent neurons and edges play the part of synaptic interaction \cite{Amit,Torres_rev,Sporns}. Experimental neuroscience has shown that enhanced electric activity induces synaptic growth and dendritic arborization
\cite{Lee,Frank,Klintsova,DeRoo}. Since the activity of a
neuron depends on the net current received from its neighbours, which is higher
the more neighbours it has, we can see node degree as a proxy for this
activity -- accounting for the local term alluded to above.
On the other hand, synaptic growth and death also depend on concentrations of various molecules, which can diffuse through large areas of tissue and therefore cannot in general be considered local. A feature of brain development in many animals is \textit{synaptic pruning} -- the large reduction in synaptic density undergone throughout infancy.
Chechik \textit{et al.} \cite{Chechik,Chechik_2} have shown that via an elimination of less needed synapses, this can reduce the energy consumed by the brain (which in a human at rest can account
for a quarter of total energy used) while maintaining near optimal memory performance. Going on this, we will take the mean degree of the network -- or mean synaptic density -- to reflect total energy consumption, hence the global terms in our attachment/detachment rules.

An alternative approach would be to consider some kind of model neurons explicitely and couple the probabilities of synaptic growth and death to neuronal dynamic variables, such as local and global fields. In a Hopfield network, for example, the expected value of the field (total incoming current) at node $i$ is proportional to $i$'s degree \cite{Torres_influence}, the total current (energy consumption) in the network therefore being proportional to the mean degree; qualitatively, these observations are likely to hold also in more realistic situations \cite{Magistretti}, although relations need not be linear. Co-evolving networks of this sort are currently attracting attention, with dynamics such as Prisoner's Dilemma \cite{Julia}, Voter Model \cite{Eguiluz_coevolution} or Random Walkers \cite{Antiqueira}. Although we consider this line of work particularly interesting, for generality and analytical tractability we opt here to use only topological information
for the attachment/detachment rules,
although our results can be applied to any situation in which the dynamical states of the elements at the nodes can be functionally related to degrees\footnote{For instance, the stationary distribution of walkers used for edge dynamics in \cite{Antiqueira} is actually obtained purely from topological information, although it can only be written in terms of local degrees for undirected networks.}.

Following a brief general analysis, we show how appropriate choices of functions induce the system to evolve towards heterogeneous (sometimes scale-free) networks while undergoing synaptic pruning in quantitative agreement with experiments. At the same time, degree-degree correlations emerge naturally, thus making the resulting networks {\it disassortative} -- as tends to be the case for most biological networks -- and leading to realistic small-world parameters.

\section{Basic considerations}
\label{sec_basic}
Consider a simple undirected network with $N$ nodes defined by the adjacency matrix $\hat{a}$, the element $\hat{a}_{ij}$ representing the existence or otherwise of an edge between nodes $i$ and $j$. Each node can be characterized by its degree, $k_{i}=\sum_{j}\hat{a}_{ij}$. Initially, the degrees follow some distribution $p(k,t=0)$ with mean $\kappa(t)$. 
We wish to study the evolution of networks in which nodes can gain or lose edges according to stochastic rules which only take into account local and global information on degrees. So as to implement this in the most general way, we will assume that at every
time step, each node has a probability of gaining a new edge, $P_{i}^{\mbox{gain}},$ to a random node; and a probability of losing a randomly chosen edge, $P_{i}^{\mbox{lose}}.$ We assume these factorize as $P_{i}^{\mbox{gain}}=u(\kappa)\pi(k_{i})$ and $P_{i}^{\mbox{lose}}=d(\kappa)\sigma(k_{i})$,
where $u$, $d$, $\pi$ and $\sigma$ can be arbitrary functions, 
but impose nothing else other than normalization.

For each edge that is withdrawn from the network, two nodes decrease in degree: $i$, chosen according to $\sigma(k_{i})$, and $j$, a random neighbour of $i$'s; so there is an added effective probability of loss $k_{j}/(\kappa N)$. Similarly, for each edge placed in the network, not only $l$ chosen according to $\pi(k_{l})$ increases its degree; a random node $m$ will also gain, with the consequent effective probability $N^{-1}$ (though see\footnote{We are ignoring the small corrections that arise because $j\neq i$ and $l\neq m$, which in any case would disappear if self-connections were allowed.}).
Let us introduce the notation $\tilde{\pi}(k)\equiv\pi(k)+N^{-1}$ and $\tilde{\sigma}(k)\equiv\sigma(k)+k/(\kappa N)$. Network evolution can now be seen as a \textit{one step process} \cite{vanKampen} with transition rates $u(\kappa)\tilde{\pi}(k)$ and $d(\kappa)\tilde{\sigma}(k)$. The expected value for the increment in a given $p(k,t) $ at each time step (which we equate with a temporal derivative) defines a master equation for the degree distribution \cite{Johnson_PRE}:
\begin{eqnarray}
\frac{\mbox{d} p(k,t)}{\mbox{d} t} =u\left(  \kappa\right)
\tilde{\pi}(k-1) p(k-1)
+d\left(  \kappa\right) \tilde{\sigma}(k+1)
p(k+1)\label{eq1}\\
-\left[  u\left(  \kappa\right)   \tilde{\pi}(k)
+d\left(  \kappa\right) \tilde{\sigma}(k)
\right] p(k,t).\nonumber
\end{eqnarray}

Assuming now that $p(k,t)$ evolves towards a stationary distribution,
$p_{\mbox{\small{st}}}(k)$,
then this must necessarily satisfy detailed balance
since it is a one step process \cite{vanKampen};
i.e., the flux of probability from $k$ to $k+1$ must equal the flux from $k+1$ to $k$, for all $k$ \cite{MarroBook}. This condition (sufficient for (\ref{eq1}) to be zero) can be written as 
\begin{equation}
\frac{\partial p_{\mbox{\small{st}}}(k)}{\partial k} =\left[ \frac{u(\kappa_{\mbox{\small{st}}})}{d(\kappa_{\mbox{\small{st}}})}   \frac{\tilde{\pi}(k)}{\tilde{\sigma}(k+1)}-1\right] p_{\mbox{\small{st}}}(k),\label{eq2}%
\end{equation}
where we have
substituted a difference for a partial derivative and
$\kappa_{\mbox{\small{st}}}\equiv\sum_{k}kp_{\mbox{\small{st}}}(k)$.
Setting $\pi$ and $\sigma$ so as to be normalized to one (i.e., $\sum_{k}p(k)\pi(k)=\sum_{k}p(k)\sigma(k)=1$, $\forall t$),
which is equivalent to saying that at each time step exactly $u(\kappa)$ nodes are chosen to gain edges and $d(\kappa)$ to lose them, 
then in the stationary state we will have $u(\kappa_{\mbox{\small{st}}})=d(\kappa_{\mbox{\small{st}}})$ since the total number of edges will be conserved.
From  (\ref{eq2})\textbf{\ }we can see that $p_{\mbox{\small{st}}}(k)$ will have an extremum at some value $k_{e}$ if it satisfies $\tilde{\pi}(k_{e})=\tilde{\sigma}(k_{e}+1)$. $k_{e}$ will be a maximum (minimum) if the numerator
in  (\ref{eq2}) is smaller (greater) than the denominator for $k>k_{e}$,
and viceversa for $k<k_{e}$. Assuming, for example, that there is one and only one such
$k_{e}$, then a maximum implies a relatively homogeneous distribution, while a
minimum means $p_{\mbox{\small{st}}}(k)$ will be split in two, and therefore
highly heterogeneous.
More intuitively, if for nodes with large enough $k$ there is a higher probability of gaining edges than of losing them, the degrees of these nodes will grow indefinitely, leading to heterogeneity. If, on the other hand, highly connected nodes always lose more edges than they gain, we will obtain quite homogeneous networks. From this reasoning we can see that a particularly interesting case (which turns out to be critical) is that in which
$\pi(k)$ and $\sigma(k)$
are such that 
\begin{equation}
\tilde{\pi}(k)=\tilde{\sigma}(k)\equiv v(k),\quad\forall k.
\label{condition}
\end{equation}
According to (\ref{eq2}), condition (\ref{condition}) means that for large $k$, $\partial p_{st}(k)/\partial k\rightarrow 0$, and $p_{st}(k)$ flattens out -- as for example a power-law does.

The standard Fokker-Planck approximation for the one step process defined by (\ref{eq1}) is \cite{vanKampen}:
\begin{eqnarray}
\frac{\partial p(k,t)}{\partial t}=\frac{1}{2}\frac{\partial^{2}}{\partial k^{2}}
\left\{
\left[d(\kappa)\tilde{\sigma}(k)+u(\kappa)\tilde{\pi}(k)  \right]
p(k,t)\right\}
\nonumber\\
+
\frac{\partial}{\partial k}
\left\{\left[d(\kappa)\tilde{\sigma}(k)-u(\kappa)\tilde{\pi}(k)  \right]
p(k,t)\right\}.
\label{eq_FokkerPlanck}
\end{eqnarray}
For transition rates which meet condition (\ref{condition}), (\ref{eq_FokkerPlanck}) can be written as:
\begin{eqnarray}
\frac{\partial p(k,t)}{\partial t}   =\frac{1}{2}\left[ d\left(  \kappa\right)+u\left(  \kappa\right)  \right]  \frac{\partial^{2}}{\partial k^{2}}\left[
v(k)p(k,t)\right]  \nonumber\\
 +\left[d\left(  \kappa\right)-  u\left(  \kappa\right)  \right]
\frac{\partial}{\partial k}\left[  v(k)p(k,t)\right].
\label{diff_drift}
\end{eqnarray}
Ignoring boundary conditions, the stationary solution must satisfy, on the one
hand, $v(k)p_{\mbox{\small{st}}}(k)=Ak+B,$ so that the diffusion is
stationary, and, on the other, $u(\kappa_{\mbox{\small{st}}})=d(\kappa
_{\mbox{\small{st}}}),$ to cancel out the drift. For this situation to be
reachable from any initial condition, $u(\kappa)$ and $d(\kappa)$ must be
monotonous functions, decreasing and increasing respectively.

\section{Synaptic pruning}

\begin{figure}
[th!]
\begin{center}
\includegraphics[
scale=0.6
]%
{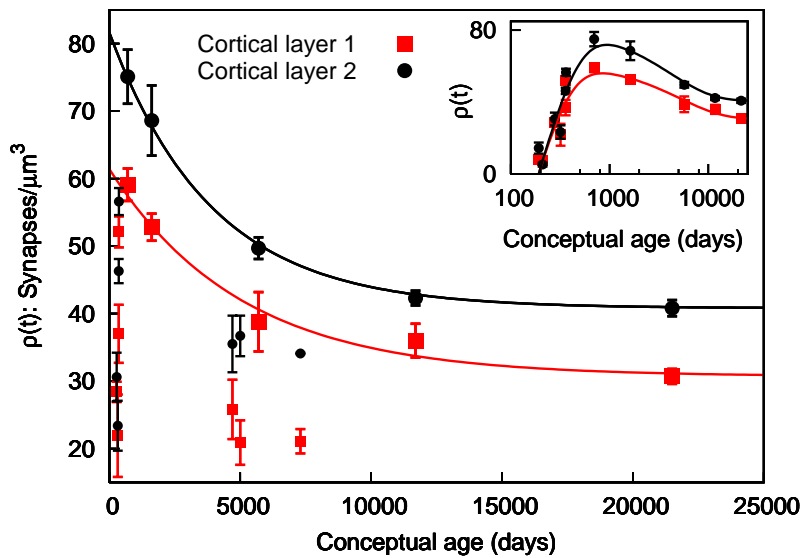}%
\caption{
Synaptic densities in layers 1 (red
squares) and 2 (black circles) of the human auditory cortex against time from conception. Data from \cite{Huttenlocher}, obtained by directly counting synapses in
tissues from autopsies. Lines follow best fits to  (\ref{rho_t}), where the parameters were: for layer 1, $\tau_{\mbox{p}}=5041$ days; and for layer 2, $\tau_{\mbox{p}}=3898$ days (for $\rho_{\mbox{f}}$ we have used the last data pints: $30.7$ and $40.8$ synapses/$\mu m^{3}$, for layers 1 and 2 respectively).
Data pertaining to the first year and to days $4700$, $5000$
$7300$, shown with smaller symbols, where omitted from the fit. Assuming the existence of transient growth factors, we can include the data points for the first year in the fit by using  (\ref{rho_t2}). This is done in the inset (where time is displayed logarithmically). The best fits were: for layer 1, $\tau_{\mbox{g}}=151.0$ and $\tau_{\mbox{p}}=5221$; and for layer 2, $\tau_{\mbox{g}}=191.1$ and $\tau_{\mbox{p}}=4184$, all in days (we have approximated $t_{0}$ to the time of the first data points, $192$ days).
}%
\label{fig_exp}%
\end{center}
\end{figure}

As a simple example,
we will first consider global probabilities which have the linear forms:
\begin{equation}
u[\kappa(t)]=\frac{n}{N}\left(  1-\frac{\kappa(t)}{\kappa_{\mbox{\small{max}}}%
}\right) \quad \mbox{ and }\quad d[\kappa(t)]=\frac{n}{N}\frac{\kappa(t)}{\kappa
_{\mbox{\small{max}}}},\label{u_d_linear}%
\end{equation}
where $n$ is the expected value of the number of additions and
deletions of edges per time step, and $\kappa_{\mbox{\small{max}}}$ is the maximum
value the mean degree can have. This choice describes a situation in which the
higher the density of synapses, the less likely new ones are to
sprout and the more likely existing ones
are to atrophy -- a situation that might arise, for instance, in the presence of a finite quantity of nutrients. Again taking into account that $\pi$ and $\sigma$
are normalized to one,
summing over $P_{i}^{\mbox{gain}}-P_{i}%
^{\mbox{lose}}$ we find that the increment in $\kappa(t)$ is $\rmd\kappa
(t)/\rmd t=2\lbrace u[\kappa(t)]-d[\kappa(t)]\rbrace=2(n/N)\left[  1-2\kappa(t)/\kappa_{\mbox{\small{max}}}%
\right] 
$
(independently of the local probabilities). 
Therefore, the mean degree will increase
or decrease exponentially with time, from $\kappa(0)$ to $\frac{1}{2}\kappa_{\mbox{\small{max}}}$.
Assuming that the initial condition is, say, $\kappa(0)=\kappa
_{\mbox{\small{max}}}$, and expressing the solution in terms of the
\textit{mean synaptic density} -- i.e., $\rho(t)\equiv\kappa(t)N/(2V),$ with $V$
the total volume considered -- we have
\begin{equation}
\rho(t)=\rho_{\mbox{f}}\left(1+e^{-t/\tau_{\mbox{p}}}\right)  ,\label{rho_t}%
\end{equation}
where we have defined $\rho_{\mbox{f}}\equiv\rho(t\rightarrow\infty)$ and the time constant for pruning is $\tau_{\mbox{p}}=\rho_{\mbox{f}}N/n$. This equation was fitted in figure \ref{fig_exp}
to experimental data on layers 1 and 2 of the human auditory cortex\footnote{Data points for three particular days (smaller symbols) are omitted from the fit, since we believe these must be from subjects with inherently lower synaptic density.} obtained during autopsies by Huttenlocher and Dabholkar \cite{Huttenlocher}. 

It seems reasonable to assume that the initial overgrowth of synapses is due to the transient existence of some kind of growth factors. If we account for these by including a nonlinear, time-dependent term $g(t)\equiv
a\exp(-t/\tau_{\mbox{g}})$ in the probability of growth, i.e., $u[\kappa(t),t]=(n/N)[1-\kappa
(t)/\kappa_{\mbox{\small{max}}}+g(t)],$ leaving $d[\kappa(t)]$ as before, we find that $\rho(t)$ becomes
\begin{equation}
\rho(t)=\rho_{\mbox{f}}\left[1+e^{-t/\tau_{\mbox{p}}}-\left( 1+e^{-t_{0}/\tau_{\mbox{p}}} \right)e^{-\frac{t-t_{0}}{\tau_{\mbox{g}}}}  \right]  ,\label{rho_t2}%
\end{equation}
where $t_{0}$ is the time at which synapses begin to form ($t=0$ corresponds to the moment of conception) and $\tau_{\mbox{g}}$ is the time constant related to growth.
The inset in figure \ref{fig_exp} shows the best fit to the auditory cortex data.
Since the contour conditions $\rho_{\mbox{f}}$ and (for  (\ref{rho_t2}))f $t_{0}$ are simply taken as the value of the last data point and the time of the first one, in each case, the time constants $\tau_{\mbox{p}}$ and $\tau_{\mbox{g}}$ are the only parameters needed for the fit.

\begin{figure}
[th!]
\begin{center}
\includegraphics[
scale=0.47
]%
{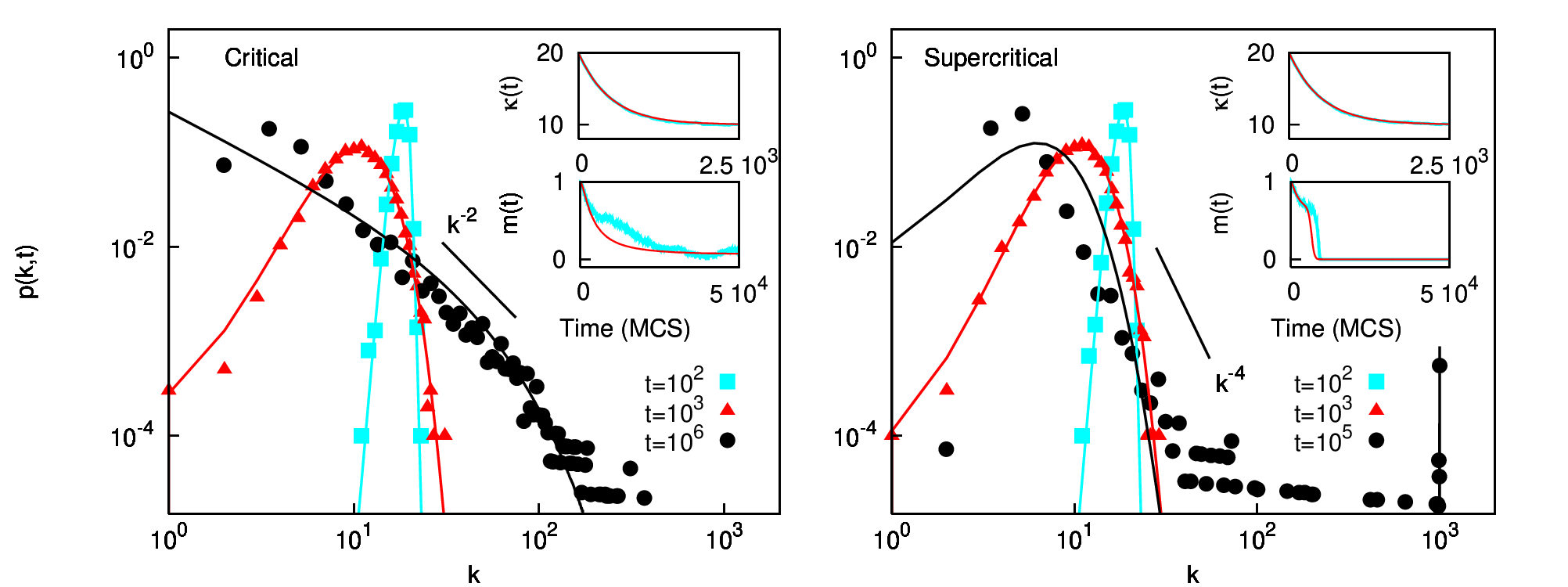}
\caption{
Evolution of the degree distributions of networks 
beginning as regular random graphs with $\kappa(0)=20$ in the critical (top)
and supercritical (bottom) regimes. Local probabilities are
$\sigma(k)=k/(\langle k\rangle N)$
in
both cases, and $\pi(k)=2\sigma(k)-N^{-1}$ and
$\pi(k)= k^{3/2}/(\langle k^{3/2}\rangle N)$
for the critical and supercritical ones, respectively. Global probabilities as in (\ref{u_d_linear}), with $n=10$ and $\kappa_{\mbox{\small{max}}}=20$. Symbols
in the main panels correspond to $p(k,t)$ at different times as obtained from
MC simulations. Lines result from numerical integration of  (\ref{eq1}). 
Insets show typical time series of $\kappa$ and $m$. Light blue lines are from MC simulations and red lines are theoretical, 
given by  (\ref{rho_t}) and (\ref{eq1}), respectively. $N=1000$.}%
\label{fig_ab}%
\end{center}
\end{figure}

\section{Phase transitions}

The drift-like evolution of the mean degree we have just illustrated with
the example of synaptic pruning is independent of the local probabilities $\pi(k)$ and
$\sigma(k)$. The effect of these is rather in the diffusive behaviour which can
lead, as mentioned, either to homogeneous or to heterogeneous final states. A useful
bounded order parameter to characterize these phases is therefore $m\equiv\exp\left(
-\sigma^{2}/\kappa^{2}\right),$ where $\sigma^{2}=\langle k^{2}%
\rangle-\kappa^{2}$ is the variance of the degree distribution ($\langle\cdot\rangle\equiv N^{-1}\sum_{i}(\cdot)$ represents an average over nodes). We will use
$m_{\mbox{\small{st}}}\equiv\lim_{t\rightarrow\infty}m(t)$ to distinguish
between the different phases, since $m_{\mbox{\small{st}}}=1$ for a regular
network and $m_{\mbox{\small{st}}}\rightarrow0$ for one following a highly
heterogeneous distribution. Although there are particular choices of probabilities which lead to  (\ref{diff_drift}), these are not the only critical cases, since the transition from homogeneous to heterogeneous stationary states can come about also with functions which never meet condition (\ref{condition}). Rather, this is a classic topological phase transition, the nature of which depends on the choice of functions \cite{Park_equilibrium,Burda,Derenyi} .

\begin{figure}
[th!]
\begin{center}
\includegraphics[
scale=0.6]%
{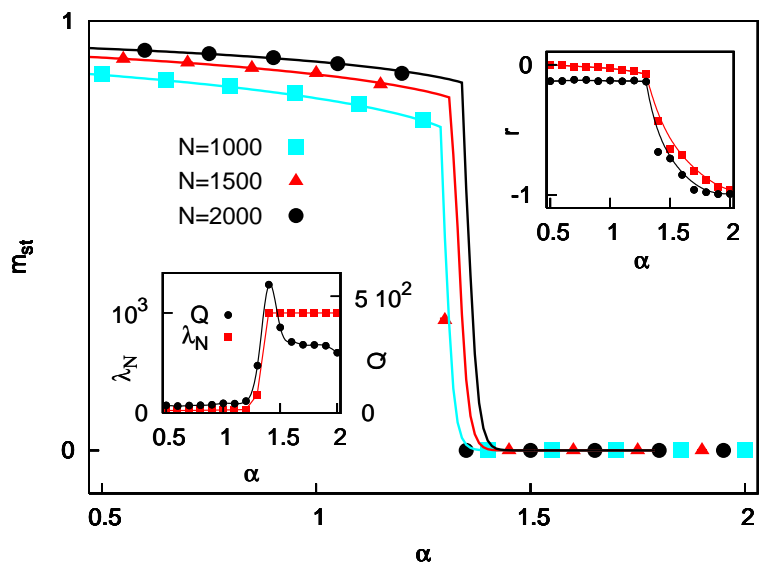}
\caption{
Phase transitions in $m_{\mbox{\small{st}}}$ for
$\pi(k)\sim k^{\alpha}$ and $\sigma(k)\sim k$, and $u(\kappa)$ and $d(\kappa)$
as in  (\ref{u_d_linear}). $N=1000$ (blue squares), $1500$ (red triangles)
and $2000$ (black circles); $\kappa(0)=\kappa_{\mbox{\small{max}}}=2n=N/50$.
Corresponding lines are from numerical integration of  (\ref{eq1}). 
The bottom left inset shows values of the highest eigenvalue of the Laplacian matrix (red squares) and of $Q=\lambda_{N}/\lambda_{2}$ (black circles), a measure of unsynchronizability; $N=1000$. The top right inset shows transitions for the same parameters in the final values of Pearson's correlation coefficient $r$ (see section \ref{sec_correlations}), both for only one edge allowed per pair of nodes (red squares) and without this restriction (black circles).
}%
\label{fig_fss}%
\end{center}
\end{figure}

Evolution of the degree distribution is shown in figure \ref{fig_ab} for
critical and supercritical choices for the probabilities, as given
by MC simulations (starting from regular random graphs) and contrasted with theory. (The subcritical regime is not shown since the stationary state has a distribution similar to the ones at $t=10^{3}$ MCS in the other regimes.) The disparity between the theory and the
simulations for the final distributions is due to the build up of certain
correlations not taken into account in our analysis. This is because the
existence of some very highly connected nodes reduces the probability of there being very
low degree nodes. In particular, if there are, say, $x$ nodes connected to the
rest of the network, then a natural cutoff, $k_{min}=x$, emerges. Note that
this occurs only when we restrict ourselves to simple networks, i.e., with
only one edge allowed for each pair of nodes. 
This topological phase transition is
shown in figure \ref{fig_fss}, where $m_{\mbox{st}}$ is plotted against parameter $\alpha$ for global probabilities as in  (\ref{u_d_linear}) and local ones $\pi(k)\sim k^{\alpha}$ and $\sigma(k)\sim k$. This situation corresponds to one in which edges are eliminated randomly while nodes have a power-law probability of sprouting new ones (note that power-laws are good descriptions of a variety of monotonous response functions, yet only require one parameter).
Although, to our knowledge, there are not yet enough empirical data to ascertain what degree distribution the structural topology of the human brain follows, it is worth noting that its functional topology, at the level of brain areas, has been found to be scale-free with an exponent very close to $2$ \cite{Eguiluz}.

In general, most other measures can be expected to undergo a transition along with its variance. For instance, highly heterogeneous networks (such as scale-free ones) exhibit the small-world property, characterized by a high \textit{clustering coefficient,}
$C\gg \langle k\rangle /N$, and a low \textit{mean minimum path,} $l\sim \ln(N)$ \cite{Watts}. A particularly interesting topological feature of a network is its {\it synchronizability} -- i.e., 
given a set of oscillators placed at the nodes and coupled via the edges, how wide a range of coupling strengths will result in them all becoming synchronized.
Barahona and Pecora showed analytically that, for linear oscillators, a network is more
synchronizable the lower the relation $Q=\lambda_{N}/\lambda_{2}$ -- where $\lambda_{N}$ and $\lambda_{2}$ are the highest and lowest non-zero eigenvalues of the Laplacian matrix ($\hat{\Lambda}_{ij}\equiv \delta_{ij}k_{i}-\hat{a}_{ij}$), respectively \cite{Barahona}.
The bottom left inset in figure \ref{fig_fss} displays the values of $Q$ and $\lambda_{N}$ obtained for the different stationary states. There is a peak in $Q$
at the critical point.
It has been argued that this tendency of heterogeneous topologies to be particularly unsynchronizable poses a paradox given the wide prevalence of scale-free networks in nature, a problem that has been deftly got around by considering appropriate weighting schemes for the edges \cite{Adilson_paradox,Chavez} (see also\footnote{Using pacemaker nodes, scale-free networks have also been shown to emerge via rules which maximize synchrony \cite{Plos}.}, and \cite{Arenas_rev} for a review). However, there is no generic reason why high synchronizability should always be desirable. In fact, it has recently been shown that heterogeneity can improve the dynamical performance of model neural networks precisely because the fixed points are easily destabilised \cite{Johnson_EPL} (as well as
conferring robustness to thermal fluctuations and 
improving storage capacity \cite{Torres_influence}). This makes intuitive sense, since, presumably,
one would not usually want all the neurons in one's brain to be doing exactly the same thing. Therefore,
this point of maximum \textit{unsynchronizability} at the phase transition may be a particularly advantageous one.

On the whole, we find that three classes of network -- homogeneous, scale-free (at the critical point) and ones composed of starlike structures, with a great many small-degree nodes connected to a few hubs -- can emerge for any kind of attachment/detachment rules. It follows that a network subject to some sort of optimising mechanism, such as Natural Selection for the case of living systems, could thus evolve towards whichever topology best suits its requirements by tuning these microscopic actions.

\section{Correlations}
\label{sec_correlations}
Most real networks have been found to exhibit degree-degree correlations, also known as {\it mixing} by degree \cite{Pastor-Satorras,Newman_rev}. They can thus be classified as {\it assortative}, when the degree of a typical node is positively correlated with that of its neighbours, or {\it disassortative,} when the correlation is negative. This property has important implications for network characteristics such as connectedness and robustness \cite{Newman_mixing_PRL,Newman_mixing_PRE}. A useful measure of this phenomenon is Pearson's correlation coefficient applied to the edges \cite{Newman_rev, Newman_mixing_PRE, Boccaletti}: $
r= ([ k_{l}k'_{l}]-[ k_{l}]^{2})/([ k_{l}^{2}]-[ k_{l}]^{2}), $
where $k_{l}$ and $k'_{l}$ are the degrees of each of the two nodes
pertaining to edge $l$, and $[\cdot]\equiv(\langle k\rangle
N)^{-1}\sum_{l}(\cdot)$ represents an average over edges; $|r|\leq 1$. Writing
$\sum_{l}(\cdot)=\sum_{ij}\hat{a}_{ij}(\cdot)$, $r$ can be expressed in
terms of averages over nodes:
\begin{equation}
  r=\frac{\langle k\rangle \langle k^{2} k_{nn}(k)\rangle - 
    \langle k^{2}\rangle^{2} }{\langle k\rangle \langle k^{3}\rangle 
    - \langle k^{2}\rangle^{2}},
  \label{r_gen}
\end{equation}
where $k_{nn}(k)$ is the mean nearest-neighbour-degree function; i.e., if $k_{nn,i}\equiv k_{i}^{-1}\sum_{j}\hat{a}_{ij}k_{j}$ is the mean degree of the neighbours of node $i$, $k_{nn}(k)$ is its average over all nodes such that $k_{i}=k$. Whereas most social networks are assortative ($r>0$) -- due, probably, to mechanisms such as homophily \cite{Newman_rev} -- almost all other networks, whether biological, technological or information-related, seem to be generically disassortative. The top right inset in figure \ref{fig_fss} displays the stationary value of $r$ obtained in the same networks as in the main panel and lower inset. It turns out that the heterogeneous regime is disassortative, the more so the larger $\alpha$. (Note that a completely homogeneous network cannot have degree-degree correlations, since all degrees are the same.) It is known that the restriction of having at most one edge per pair of nodes induces disassortativity \cite{Park_correlations,Maslov}. However, in our case this is not the sole origin of the correlations, as can also be seen in the same inset of figure \ref{fig_fss}, where we have plotted $r$ for networks in which we have lifted the restriction and allowed any number of edges per pair of nodes. In fact, when multiple edges are allowed, the correlations are slightly stronger.

To understand how these correlations come about, consider a pair of nodes $(i,j)$, which, at a given moment, can either be occupied by an edge or unoccupied. We will call the expected times of permanence for occupied and unoccupied states $\tau_{ij}^{\mbox{o}}$ and $\tau_{ij}^{\mbox{u}}$, respectively. After sufficient evolution time (so that occupancy becomes independent of the initial state\footnote{Note that this will always happen eventually since the process is ergodic.}), the expected value of the corresponding element of the adjacency matrix, $E(\hat{a}_{ij})\equiv \hat{\epsilon}_{ij}$, will be
$$
\hat{\epsilon}_{ij}=\frac{\tau_{ij}^{\mbox{o}}}{\tau_{ij}^{\mbox{o}}+\tau_{ij}^{\mbox{u}}}.
$$
If $p_{ij}^{+}$ ($p_{ij}^{-}$) is the probability that $(i,j)$ will become occupied (unoccupied) given that it is unoccupied (occupied), then $\tau_{ij}^{\mbox{o}}\sim 1/p_{ij}^{-}$ and $\tau_{ij}^{\mbox{u}}\sim 1/p_{ij}^{+}$, yielding
$$
\hat{\epsilon}_{ij}=\left(1+\frac{p_{ij}^{-}}{p_{ij}^{+}}\right)^{-1}.
$$
Taking into account the probability that each node has of gaining or losing an edge, we obtain\footnote{Again, we are ignoring corrections due to the fact that $i$ is necessarily different from $j$.}:
$p_{ij}^{+}=u(\langle k\rangle)N^{-1}[\pi(k_{i})+\pi(k_{j})]$ and $p_{ij}^{-}=d(\langle k\rangle)[\sigma(k_{i})/k_{i}+\sigma(k_{j})/k_{j}]$.
Then, assuming that the network is sparse enough that $p_{ij}^{-}\gg p_{ij}^{+}$ (since the number of edges is much smaller than the number of pairs), and particularising for power-law local probabilities $\pi(k)\sim k^{\alpha}$ and $\sigma(k)\sim k^{\beta}$, the expected occupancy of the pair is
$$
\hat{\epsilon}_{ij}\simeq\frac{p_{ij}^{+}}{p_{ij}^{-}}=\frac{u(\langle k\rangle)}{d(\langle k\rangle)}\frac{\langle k^{\beta}\rangle}{\langle k^{\alpha}\rangle N}
\left(\frac{k_{i}^{\alpha}+k_{j}^{\alpha}}{k_{i}^{\beta-1}+k_{j}^{\beta-1}} \right).
$$
Considering the stationary state, when $u(\langle k\rangle)=d(\langle k\rangle)$, and for the case of random deletion of edges, $\beta=1$ (so that the only nonlinearity is due to $\alpha$), the previous expression reduces to
\begin{equation}
\hat{\epsilon}_{ij}\simeq\frac{\langle k\rangle}{2\langle k^{\alpha}\rangle N}
\left(k_{i}^{\alpha}+k_{j}^{\alpha} \right).
\label{epsi_b1}
\end{equation}
(Note that this matrix is not consistent term by term, since $\sum_{j}\hat{\epsilon}_{ij}\neq k_{i}$, although it is globally consinstent: $\sum_{ij}\hat{\epsilon}_{ij}=\langle k\rangle N$.)
The nearest-neighbour-degree function is now
$$
k_{nn}(k_{i})=\frac{1}{k_{i}}\sum_{j}\hat{\epsilon}_{ij}k_{j}=
\frac{\langle k\rangle}{2\langle k^{\alpha}\rangle}
(\langle k\rangle k_{i}^{\alpha-1}+\langle k^{\alpha+1}\rangle k_{i}^{-1})
\label{knn}
$$
(a decreasing function for any $\alpha$),
with the result that Pearson's coefficient becomes
\begin{equation}
r=\frac{1}{\langle k^{\alpha}\rangle}\left(
\frac{\langle k\rangle^{3}\langle k^{\alpha+1}\rangle-\langle k^{2}\rangle^{2}\langle k^{\alpha}\rangle }{\langle k\rangle \langle k^{3}\rangle -\langle k^{2}\rangle^{2}}
\right).
\label{eq_r}
\end{equation}

More generally, one can understand the emergence of these correlations in the following way. For the network to become heterogeneous, we must have $\pi(k)+N^{-1}\geq\sigma(k)+k/(\langle k\rangle N)$ for large enough $k$, so that highly connected nodes do not lose more edges than they can acquire (see section \ref{sec_basic}). This implies that $\pi(k)$ must be increasing and approximately linear or superlinear. The expected value of the degree of a node $i$, chosen according to $\pi(k_{i})$, is then $E(k_{i})=N^{-1}\sum_{k} \pi(k)k\gtrsim \langle k^{2}\rangle/\langle k\rangle$, while that of its new, randomly chosen neighbour, $j$, is only $E(k_{j})=\langle k\rangle$. This induces disassortative correlations which can never be compensated by the breaking of edges between nodes whose expected degree values are $N^{-1}\sum_{k} \sigma(k)k$ and $\langle k^{2}\rangle/\langle k\rangle$ if $\sigma(k)$ is an increasing function. It thus ensues that a scenario such as the one analysed in this paper will never lead to assortative networks except for some cases in which $\sigma(k)$ is a decreasing function -- meaning that less connected nodes should be more likely to lose edges. Assortativity could, however, arise if there were some bias also on the node chosen to be $i$'s neighbour, e.g. on the postsynaptic neuron -- which is precisely what happens in most social networks, where individuals do not generally choose their friends, partners, etc. randomly. Although there seem to be other reasons for the ubiquity of disassortative networks in nature \cite{Johnson_PRL}, it is possible that the generality of the scenario studied here may also play a part.

We can use the expected value matrix $\hat{\epsilon}$ to estimate other magnitudes. For example, the clustering coefficient, as defined by Watts and Strogatz \cite{Watts}, is an average over nodes of $C_{i}$, with $C_{i}$ the proportion of $i$'s neighbours which are connected to each other; so its expected value is $E(C_{i})=\hat{\epsilon}_{jl}$ conditioned to $j$ and $l$ being neighbours of $i$'s. This means that, on average, we can make the approximation that $k_{j}=k_{l}=\langle k_{nn}\rangle=\langle k\rangle[\langle k\rangle\langle k^{\alpha-1}\rangle+\langle k^{\alpha+1}\rangle\langle k^{-1}\rangle]/(2\langle k^{\alpha}\rangle)$. Substituting this value in  (\ref{epsi_b1}), and taking into account that one edge of $j$'s and one of $l$'s are taken up by $i$, we have
\begin{equation}
C\simeq \frac{\langle k\rangle}{\langle k^{\alpha}\rangle N}(\langle k_{nn}\rangle-1)^{\alpha}.
\label{C}
\end{equation}
For a rough estimate of the mean minimum path (the minimum path between two nodes being the smallest number of edges one has to follow to get from one to the other), we can procede as in \cite{Albert_Rev}. For a given node, let us define the number of nearest neighbours, $z_{1}$, next-nearest neighbours, $z_{2}$, and in general $m$th neighbours, $z_{m}$. Using the relation
$
z_{m}=z_{1}\left(z_{2}/z_{1}\right)^{m-1},
$
and assuming that the network is connected and can be obtained in $l$ steps, this yields
\begin{equation}
1+\sum_{1}^{l}z_{m}=N.
\label{lz}
\end{equation}
On average, $z_{1}=\langle k\rangle$ and $z_{2}=\langle k\rangle[(1-C)\langle k_{nn}\rangle-1]$ (since for each second nearest neighbour, one edge goes to the reference node and a proportion $C$ to mutual neighbours). Now, if $N\gg z_{1}$ and $z_{2}\gg z_{1}$,  (\ref{lz}) leads to
\begin{equation}
l\simeq 1+\frac{\ln(N/\langle k\rangle)}{\ln[(1-C)\langle k_{nn}\rangle-1]}.
\label{l}
\end{equation}

\section{The C. Elegans neural network}

There exists a biological neural network which has been entirely mapped (although not, to the best of our knowledge, at different stages of development) -- that of the much-investigated worm \textit{C. Elegans} \cite{White, Watts}. With a view to testing whether such a network could arise via simple stochastic rules of the kind we are here considering, we ran simulations for the same number of nodes, $N=307$, and (stationary) mean degree,
$\langle k\rangle =14.0$ (in the simple, undirected representation of the network).
Using the global probabilities given by  (\ref{u_d_linear}) and local ones $\pi(k)\sim k^{\alpha}$ and $\sigma(k)\sim k$ (as in figure \ref{fig_fss}), we obtain a surprising result. Precisely at the critical point, $\alpha=\alpha_{c}\simeq1.35$, there are some remarkable similarities between the biological network and the ones produced by the model.

\begin{figure}
[th!]
\begin{center}
\includegraphics[
scale=0.6
]%
{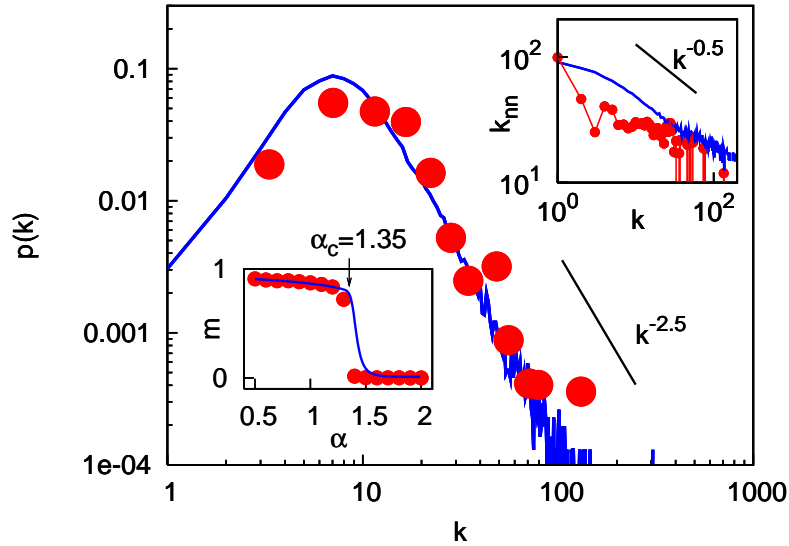}
\caption{
Degree distribution (binned) of the {\it C. Elegans} neural network (circles) \cite{White} and that obtained with MC simulations (line) in the stationary state ($t=10^{5}$ steps) for an equivalent network in which edges are removed randomly ($\beta =1$) at the critical point ($\alpha=1.35$). $N=307$, $\kappa_{\mbox{st}}=14.0$, averages over $100$ runs. Global probabilities as in (\ref{u_d_linear}). The slope is for $k^{-5/2}$. Top right inset: mean-neighbour-degree function $k_{nn}(k)$ as measured in the same empirical network (circles) and as given by the same simulations (line) as in the main panel. The slope is for $k^{-1/2}$. Bottom left inset: $m_{\mbox{st}}$ of equivalent network for a range of $\alpha$, both from simulations (circles) and as obtained with  (\ref{eq1}). (See also table \ref{table_CElegans}.)
}%
\label{fig_CElegans}%
\end{center}
\end{figure}

Figure \ref{fig_CElegans} displays the degree distributions, both for the empirical network and for the average (stationary) simulated network corresponding to the critical point,
while the top inset shows the mean-nearest-neighbour degree function $k_{nn}(k)$ for the same networks.
Both $p(k)$ and $k_{nn}(k)$ of the simulated networks can be seen to be very similar to those measured in the biological one. 
Furthermore, as is displayed in table \ref{table_CElegans}, the clustering coefficient obtained in simulation is almost the same as the empirical one. The mean minimum path is similar though slightly smaller in simulation, probably due to the worm's brain having modules related to functions \cite{Arenas_celegans}.
Finally,
Pearson's coefficient is
also in fairly good agreement, although the simulated networks are actually a bit more disassortative. It should, however, be stressed that the simulation results are for averages over $100$ runs, while the biological system is equivalent to a single run; given the small number of neurons, statistical fluctuations can be fairly large, so one should refrain from attributing too much importance to the precise values obtained -- at least until we can average over $100$ worms.
Table \ref{table_CElegans} also shows the values of $C$, $l$ and $r$ both as estimated form the theory laid out in section \ref{sec_correlations}, and for the equivalent network in the \textit{configuration model} \cite{Newman_rev} -- generally taken as the null model for heterogeneous networks, where the probability of an edge existing between nodes $i$ and $j$ is $k_{i}k_{j}/(\langle k\rangle N)$. It is clear that whereas the configuration-model predictions deviate substantially from the
magnitudes measured in the C. Elegans neural network, the growth process we are here considering accounts for them quite well.
\begin{table}
[th!]
\begin{center}
\caption{\label{table_CElegans}Values of small-world parameters $C$ and $l$, and Pearson's correlation coefficient $r$, as measured in the neural network of the worm {\it C. Elegans} \cite{White}, and as obtained from simulations in the stationary state ($t=10^{5}$ steps) for an equivalent network at the critical point when edges are removed randomly -- i.e., for $\alpha=1.35$ and $\beta=1$. $N=307$, $\kappa_{\mbox{st}}=14.0$; averages over $100$ runs and global probabilities as in  (\ref{u_d_linear}). Theoretical estimates correspond to  (\ref{C}), (\ref{l}) and (\ref{eq_r}) applied to the networks generated by the same simulations. The last column lists the respective \textit{configuration model} values: $C$ and $l$ are obtained theoretically as in \cite{Newman_rev}, while $r$, from MC simulations as in \cite{Maslov}, is the value expected due to the absence of multiple edges. (See also figure \ref{fig_CElegans}.)}
\begin{tabular}{@{}lllll}
\br
\cr &Experiment &Simulation &Theory &Config.\\
\mr
$C$ & 0.28 & 0.28 & 0.23 & 0.15 \\ 
$l$ & 2.46 & 2.19 &  1.86 & 1.96\\
$r$ & -0.163 & -0.207 & -0.305 & -0.101\\
\br
\end{tabular}
\end{center}
\end{table}
It is interesting that it should be at the critical point that a structural topology so similar to the empirical one emerges, since it seems that the brain's functional topology may also be related to a critical point \cite{Chialvo_critical,Chialvo_AIP}.

\section{Discussion}

With this work we have attempted, on the one hand, to extend our understanding of evolving networks so that any choice of transition probabilities dependent on local and/or global degrees can be treated analytically, thereby obtaining some model-independent results; and on the other, to illustrate how such a framework can be applied to realistic biological scenarios. For the latter, we have used two examples relating to rather different nervous systems:
\\
{\bf i)} synaptic pruning in humans, for which the use of nonlinear global probabilities reproduces the initial increase and subsequent depletion in synaptic density in good accord with experiments -- to the extent that nonmonotonic data points spanning a lifetime can be very well fitted with only two parameters; and
\\
{\bf ii)} the structure of the {\it C. Elegans} neural network, for which it turns out that by only considering the numbers of nodes and edges, and imposing random deletion of edges and power-law probability of growth, the critical point leads to networks exhibiting many of the worm's nontrivial features -- such as the degree distribution, small-world parameters, and even level of disassortativity.

These examples indicate that it is not farfetched to contemplate how many structural features of the brain or other networks -- and not just the degree distributions -- could arise by simple stochastic rules like the ones considered; although, undoubtedly, other ingredients such as natural modularity \cite{Arenas_celegans}, a metric \cite{Kaiser_spatial} or functional requirements \cite{Sporns} can also be expected to play a role in many instances.  We hope, therefore, that the framework laid out here -- in which for simplicity we have assumed the network to be undirected and to have a fixed size, although generalizations are straightforward -- may prove useful for interpreting data from a variety of fields. It would be particularly interesting to try to locate and quantify the biological mechanisms assumed to be behind this kind of network dynamics.

\section*{Acknowledgements}

We are very grateful to Dante R. Chialvo and to Alex Arenas for critical readings of the manuscript and fruitful suggestions, as well as to V\'{\i}ctor M. Egu\'{\i}luz, Miguel A. Mu{\~n}oz, Pedro Garrido, Jorge F. Mejias and Sebastiano de Franciscis for useful discussions.
This work was supported by Junta de Andaluc{\'i}a project FQM--01505 and by Spanish MEC--FEDER project FIS2009--08451. 




\end{document}